\documentclass[aps,prb,twocolumn,amsmath,amssymb,superscriptaddress]{revtex4}

\usepackage{amssymb}
\usepackage{amsmath}
\usepackage{graphicx}
\usepackage{textcomp}
\usepackage{color}
\usepackage{amsfonts}
\usepackage{epsfig}
\usepackage{hyperref}
\usepackage{bm}
\usepackage[caption=false]{subfig} 
\usepackage{booktabs}
\usepackage{tabularx}
\usepackage{array}

\newcommand{\arctanh}[1]{\operatorname{arctan}}

\newcolumntype{M}[1]{>{\centering\arraybackslash}m{#1}}

\DeclareMathAlphabet\mathbfcal{OMS}{cmsy}{b}{n}

\begin{document}

\title{Current-induced spin polarization at metallic surfaces from first-principles}
\author{Andrea Droghetti}
\email{andrea.droghetti@tcd.ie}
\affiliation{School of Physics and CRANN, Trinity College, The University of Dublin, Dublin 2, Ireland}
\author{Ilya V. Tokatly}
\affiliation{Nano-Bio Spectroscopy Group and European Theoretical Spectroscopy Facility (ETSF), Departamento de Pol\'imeros y Materiales Avanzados: F\'isica, Qu\'imica y Tecnolog\'ia, Universidad del Pa\'is Vasco (UPV/EHU), Av. Tolosa 72, 20018 San Sebasti\'{a}n, Spain}
\affiliation{IKERBASQUE, Basque Foundation for Science, 48009 Bilbao, Spain}
\affiliation{Donostia International Physics Center (DIPC), 20018 Donostia-San Sebasti\'{a}n, Spain}
\affiliation{ITMO University, Department of Physics and Engineering, Saint-Petersburg, Russia}

\begin{abstract}
We present the results of first-principles calculations based on density functional theory estimating the magnitude of the current induced spin polarization (CISP) at the surfaces of the $5d$ transition metals with fcc and bcc crystal structures. We predict that the largest surface CISP occurs for W and Ta, whereas CISP is considerably weaker for Pt and Au surfaces. We then discuss how CISP emerges over a length scale equal to few atomic layers as opposed to the spin accumulation characteristic of the SHE, which is related to the materials' spin diffusion length. Finally, using our estimates for the CISP magnitude, we suggest that the spin density appearing near W surfaces in experiments is mostly due to CISP, whereas that at Pt surfaces stems from the Hall effect. 
  
\end{abstract}

\maketitle

Spin-charge conversion phenomena\cite{ot.sh.17, ha.ot.18} mediated by the spin-orbit coupling (SOC) have opened new promising pathways to control and detect electrons' spins for next generation spintronic devices. Prominent examples of such conversion phenomena are the current-induced spin-polarization (CISP)\cite{ga.tr.19} and its Onsager reciprocal effect. The conduction electrons of some non-magnetic materials become spin-polarized in presence of a flowing {\it dc} charge current, and, in turn, a charge current is generated as a response to a non-equilibrium spin-polarization.\\

CISP was predicted more than four decades ago\cite{iv.pi.78} and first observed in tellurium \cite{vo.iv.79}. 
Later, the phenomenon was investigated in the two-dimensional electron gas \cite{iv.ly.89,ar.ly.89,ed.90,le.na.85} and detected optically in semiconducting heterostructures \cite{ka.my.04,ka.my.04science,si.my.05,ya.he.06,ch.ch.07,no.tr.14}. 
CISP in these 2D systems is called Rashba-Edelstein effect or, equivalently, inverse spin-galvanic effect. Recently, CISP has also been reported in the semimetallic TaSi$_2$ (Ref. \onlinecite{sh.ak.21}), while the possibility to modulate the effect through electrostatic gating has been demonstrated in Te nanowires \cite{ca.su.22}. 
%Its distinctive feature is that the spin-polarization is spatially homogeneous extending over the whole sample. 
Yet, to date, CISP has mostly been studied in $4d$ and $5d$ transition metal films, where a homogeneous spin-polarization emerges at surfaces and interfaces\cite{zh.ya.14}.
%Recently CISP has been particularly studied in $4d$ and $5d$ transition metal films with the homogeneous spin-polarization emerging at surfaces and interfaces\cite{zh.ya.14}. 
When one of these $4d$ or $5d$ films is in proximity to a ferromagnet, the interfacial CISP can exert a torque on the ferromagnet's magnetization\cite{mi.mo.11,gh.ga.17,ch.li.18}. This type of spin-orbit torque\cite{ma.ze.19} was predicted in early model calculations by Manchon and Zhan\cite{ma.zh.08} %and, since the first experimental demonstration in Pt/Co\cite{mi.mo.11}, it 
and has extensively been studied as a mean to write information in magnetoresistive RAMs\cite{sh.li.21}.
%In heterostructures composed of  ferromagnetic and non-magnetic metallic thin films, CISP at the film interfaces is often claimed as the cause for the torque acting on the magnetization of the ferromagnetic layers\cite{Miron,Ghosh,XChen}. 
Along with CISP, its reciprocal effect, often called spin galvanic effect, has also been demonstrated and extensively studied in many materials. The experiments initially considered semiconductor quantum wells \cite{ga.iv.01,ga.iv.02,ga.da.06}, while, more recently, the focus has shifted towards metallic interfaces\cite{sa.vi.13,sa.de.15,is.ma.16,no.ta.15,ma.ad.17,yu.mi.20}, metal-insulator interfaces\cite{ka.ko.16,ts.ka.18}, topological insulators \cite{sh.no.14}, van der Waals heterostructures \cite{gh.ka.19} and the 2D electron gas forming at oxide interfaces \cite{le.fu.16,qi.ho.17,va.no.19}. \\
%Over the last decade, CISP has been extensively studied at materials surfaces and interfaces in thin films heterostructures due to its potential relevance for novel magneto-resistive RAMS. The spin-polarization induced at a surface of a non-magnetic film can can exert a a torque on the magnetization of an adjacent ferromagnetic layer\cite{Miron,Ghosh,XChen}.    

In spite of the large number of studies dedicated to CISP, reliable estimates of its magnitude in materials remain scarce. In device experiments, the spin-polarization is not directly measurable, and it is extrapolated from electrical signals via complex analyses and fits to effective models \cite{ho.se.12,li.va.14,sa.ph.20}. As such, the conclusions are often controversial \cite{ti.ho.19}. The problem becomes even more significant in the case of $4d$ and $5d$ transition metals and their heterostructures with ferromagnetic layers. In these systems, surface and interfacial CISP is often accompanied by the spin Hall effect (SHE)\cite{si.va.15,ho.13}. Although the SHE is a bulk phenomenon, it manifests itself at surfaces as a spin accumulation, which adds up to the CISP. In practice, separating the CISP from the spin accumulation due to the SHE is a challenging and  debated problem \cite{al.ma.15,du.ga.20,sh.fe.21,yu.li.18}. 
%In all experimental reports of spin-orbit torque, it is indeed very debated which of the two effects is dominant.
Recently, attempts to directly measure the magnitude of CISP at metal surfaces were made in spin-polarized positron beam experiments \cite{zh.ya.14,zh.ya.15}, but the reported values appear surprisingly large. \\

Given the outstanding difficulties in extracting the CISP magnitude from experiments, first-principles calculations could potentially be very helpful to get benchmark results, as recently shown for bulk Te, TaSi$_2$ (Ref. \onlinecite{ro.ce.22}), and the (111) surface of gold\cite{to.kr.15}. However, to our knowledge, there have been no calculations for the most common materials used in experiments, namely Pt and W or, in fact, for any other transition metal besides Au. The purpose of this paper is then to close this knowledge gap. \\

We use first-principles calculations based on density functional theory (DFT) to quantitatively estimate the magnitude of CISP at the surfaces of the most investigated $5d$ transition metals. We obtain the charge current and the corresponding spatial dependent spin density in thick slabs by populating the electronic bands as implied by the relaxation time approximation for electron transport. Our results indicate that the largest CISP occurs for W surfaces, and, furthermore, they provide some insight on how CISP can be  distinguished from the SHE. Specifically, we show that the two phenomena lead to a non-equilibrium spin density extending over different length scales, and we point out for what materials CISP might be more significant than the SHE. \\

CISP is described by the equation \cite{ga.tr.19}
\begin{equation}
 S^{a}=\gamma^{a}_{ b} j_{b};\,\,\,a,b=x,y,z\label{eq.SvsJ}
\end{equation}
which linearly couples the non-equilibrium spin density $\mathbf{S}=(S^x,S^y,S^z)$ to the charge current density $\mathbf{j}=(j_x,j_y,j_z)$ (note that we here employ Einstein's convention of repeated indices). The coefficients $\gamma^a_b$, which effectively describe the CISP magnitude, are material-specific. %They can only be estimated theoretically via first-principles calculations, for example, as explained in the following. 
Nonetheless, we note that the emergence of CISP in a system is completely determined by symmetry. %Any specific coefficient $\gamma^a_{b}$ in Eq. (\ref{eq.SvsJ}) will not vanish only if the corresponding components $s^{a}$ of the spin density and $j_{b}$ of the current density transform in the same way under all system symmetry operations. 
From a mathematical point of view, $\mathbf{j}$ and $\mathbf{S}$ are a polar and an axial vector, respectively. Thus, $\gamma^a_{b}$ is a component of a second-rank pseudo-tensor $\pmb{\gamma}$.
Non-zero second-rank pseudo-tensors are allowed by symmetry only in systems whose structure is characterized by a gyrotropic point symmetry group\cite{dr.to.22}. In other words, this means that CISP can be present only in gyrotropic media, first studied for their natural optical activity\cite{Landau}. Gyrotropic point groups are listed, for example, in Ref. \onlinecite{we.la.20}. They form a subset of non-centrosymmetric groups, i.e., not all non-centrosymmetric groups are gyrotropic. Thus, breaking inversion symmetry in a system is not enough to observe CISP, and the symmetry requirements are more stringent. The materials where CISP can be observed, are either chiral, polar, have $S_ 4$ or $D_{2d}$ point group. \\

%Notably, different from CISP, the SHE does not require gyrotropy. This is the first fundamental difference between the two effects.
%\cite{gyrotropic crystal point group} firstly studied because of their natural optical activity\cite{Landau}. 
%of point groups $\mathbb{O}$, $\mathbb{T}$, $\mathbb{C}_1$, $\mathbb{C}_2$, $\mathbb{C}_3$, $\mathbb{C}_4$, $\mathbb{C}_6$, $\mathbb{D}_2$, $\mathbb{D}_3$, $\mathbb{D}_4$, $\mathbb{D}_6$, $\mathbb{C}_s$, $\mathbb{C}_{2v}$, $\mathbb{C}_{3v}$, $\mathbb{S}_4$, $\mathbb{D}_{2d}$, $\mathbb{C}_{4v}$ and $\mathbb{C}_{6v}$. 
%Gyrotropic point groups, which are listed in Ref. \cite{dr.to.22}, are only a subset of non-centrosymmetric groups. 
%Notably, although CISP is often associated to the Spin Hall effect (SHE), and they are generally considered to appear together in a material\cite{Sinova}, the symmetry requirement for the two effects differ as SHE only needs broken inversion symmetry, and not gyrotropy, to emerge. It is therefore possible to find materials showing SHE, but not CISP. Furthermore, as discussed in the following, there is no quantitative correlation between the spin Hall angle and components of $\pmb{\gamma}$.\\

Transition metals have non-gyrotropic crystal structures. Hence CISP is absent as a bulk effect. However, CISP emerges at surfaces and interfaces which are locally gyrotropic. 
In the literature, the effect has often been associated to surface bands with spin textures in momentum space (e.g. Rashba-like states\cite{la.mc.96}) 
and modelled in terms of the 2D Rashba-Edelstein effect\cite{sa.vi.13}. Such description is however not complete. 
The surface states' contribution to the total surface spin density in Eq. (\ref{eq.SvsJ}) is minor \cite{to.kr.15}. 
The largest contribution originates from the continuum
of bulk states scattering off the surface \cite{to.kr.15,bo.ho.10}.
The effect is very reminiscent of the Friedel oscillations of the charge density.
For a semi-infinite jellium model, taken to be representative of metals,
the induced spin density is confined at the surface over a distance of the order of $k_F^{-1}$, where $k_F$ is the Fermi wave-vector (of the order of few \AA), and rapidly decays inside the bulk.
For real material systems, accurate predictions are only possible by means of detailed microscopic calculations, which take into account the effect of the atomic SOC on both bulk and surface electronic states. This is done in the following.\\%, where we compute the spin density of Eq. (\ref{eq.SvsJ}) using the DFT band structures and inducing a homogeneous current within the standard relaxation time approximation\cite{to.kr.15}. \\ 

The DFT calculations are performed by means of a development version of the SIESTA code \cite{so.ar.02}. We use the local spin density approximation (LSDA) for the exchange-correlation density functional\cite{ba.he.72,ba.he.72_2}. The SOC is included by means of the on-site approximation of Ref. \onlinecite{fe.ol.06}, which we have generally found accurate even for materials with complex spin textures (e.g., Ref. \onlinecite{ja.na.15}). We treat core electrons with norm-conserving Troullier-Martin pseudopotentials\cite{tr.ma.91_1,tr.ma.91_2}. The valence states are expanded through a numerical atomic orbital basis set including multiple-$\zeta$ and polarization functions.  The cutoff radii of the basis orbitals for Ta, W and Pt are obtained from Ref. \onlinecite{ri.ga.15}. The cutoff radii of the basis orbitals of Ir and Pt are the same. 
For all systems, the pseudopotentials and basis sets have been validated to closely reproduce the band structures calculated with the Quantum Espresso plane-wave code \cite{gi.ba.09}. Assuming, the relaxation time approximation\cite{to.kr.15}, the current is introduced by populating with electrons the Kohn-Sham electronic bands of energy $E_{n,\mathbf{k}}$ and momentum $\hbar\mathbf{k}$ according to a displaced Fermi distribution $f(E_{n,\mathbf{k}}-\mathbf{v}_d\mathbf{k})=[e^{-\beta(E_{n,\mathbf{k}}-E_F-\hbar\mathbf{v}_d\mathbf{k})}+1]^{-1}$, where $\mathbf{v}_d$ is the electron drift velocity.
The charge current is then evaluated through the ``bond currents'' method as explained in Refs. \onlinecite{dr.to.22,ru.dr.20}. %We set $\hbar\vert\mathbf{v}_d\vert$ of the order of $10^{-3}$ eV\AA, and 
We employ $101 \times 101$ $\mathbf{k}$-points.
SIESTA returns the spin density $\mathbf{s}(x,y,z)$ on a real space grid.
Here we use a very dense grid, specified via a mesh cutoff equal to $1000$ Ry, for accurately resolving the spin density oscillations. %The charge current is evaluated through the ``bond currents'' method as explained in Refs. \onlinecite{dr.to.22,ru.dr.20}.
We vary $\hbar\vert\mathbf{v}_d\vert$ between $10^{-3}$ and $10^{-2}$ eV\AA~ and verify that the modulus of the spin density changes linearly as a function of the current density as for Eq. (\ref{eq.SvsJ}). We consider the stable crystal structures and the experimental lattice vectors for all materials. We carry out the calculations for slabs with in-plane unit cell and whose thicknesses vary between $24$ and $46$ layers depending on the system. For all cases we check that the spin density is converged with respect to the slab thickness.\\

\begin{figure}[h]
\centering\includegraphics[width=0.48\textwidth,clip=true]{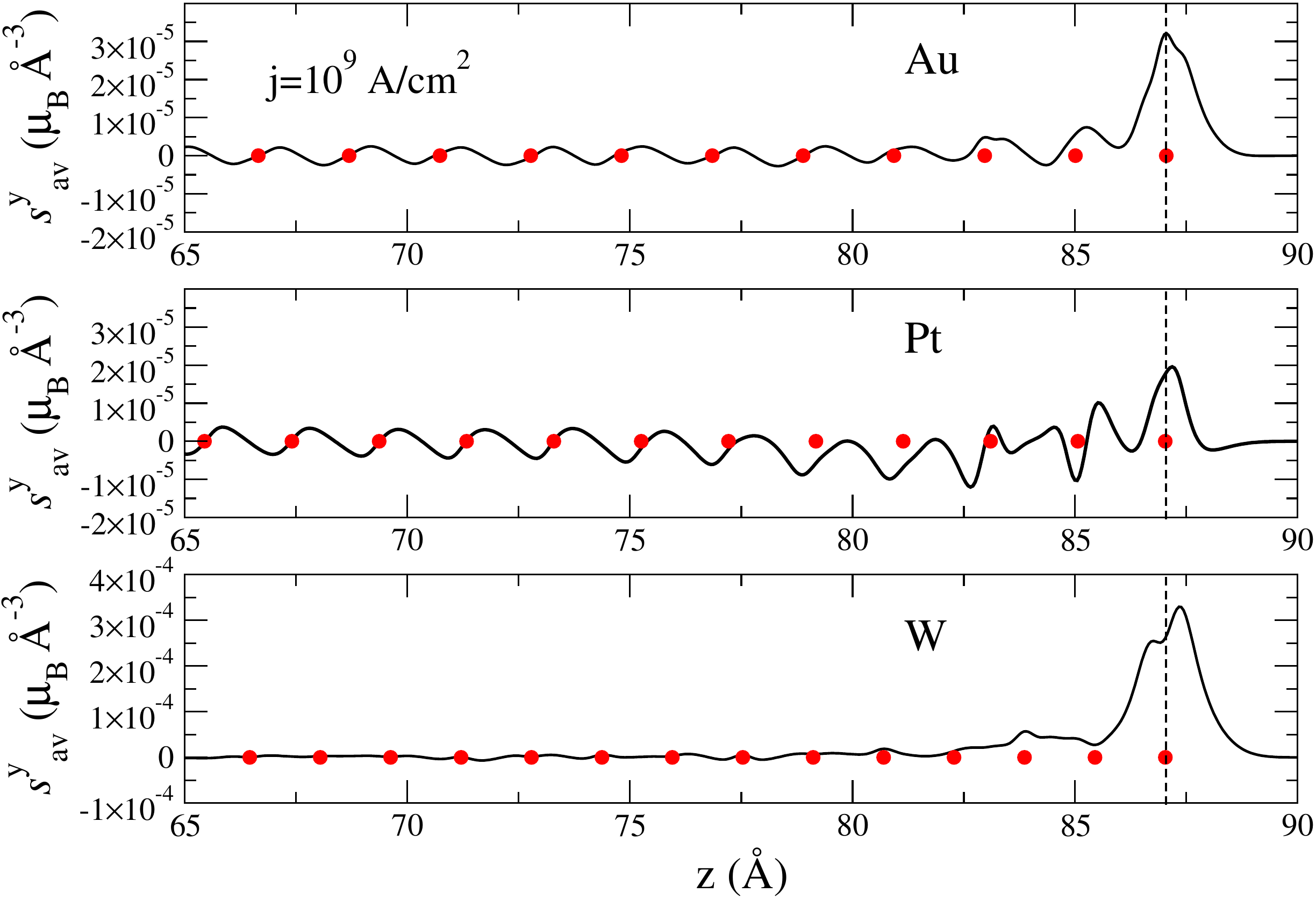}
\caption{Spin-$y$ density $s_{\mathrm{av}}^y(z)$ of Au(001), Pt(001), and W(001) calculated for an applied current density $j_x=10^9$ A/cm$^2$ along the $x$ direction parallel to the surface. The atom positions along $z$ are indicated by the red filled circles. The dashed line marks the position of the surface. Note the different scale used along the vertical axis in the three panels.}
\label{fig.m}
\end{figure}

We investigate the $5d$ transition metals with room temperature equilibrium crystal structures which are either fcc (Au, Pt, Ir) or bcc (Ta, W). 
Unless stated otherwise, we consider the (001) surfaces, which are described by the gyrotropic point group $\mathbb{C}_{4v}$. 
We assume a Cartesian frame of reference such that the $x$- and $y$-axis coincide with the (100) and (010) crystal directions. The surface normal lays along the $z$-axis, and we indicate the normal unit vector as $\mathbf{z}$. 
The only non-zero components of the pseudo-tensor $\pmb{\gamma}$ allowed by symmetry in Eq. (\ref{eq.SvsJ}), are then $\gamma^x_{y}$ and $\gamma^y_{x}$, and, additionally, $\gamma^x_{y}=\gamma^y_{x}\equiv \gamma$. Thus, Eq. (\ref{eq.SvsJ}) can be rewritten as $\mathbf{s}=\gamma\mathbf{z}\times\mathbf{j}$ (Ref. \onlinecite{to.kr.15}). %In contrast, the hcp systems exhibit anisotropy, and therefore $\gamma^x_{y}\neq\gamma^y_{x}$. 
The calculations are carried out is such a way that the charge current is along the $x$ ($y$) direction. We then obtain the in-plane average spin-$y$ (-$x$) density, also called ``spin density profile'', $s_{\mathrm{av}}^{y(x)}(z)=\int dxdy\, s^{y(x)}(x,y,z)/\mathcal{A}$, where $\mathcal{A}$ is the unit cell area in the $xy$ plane. Finally, the total surface spin density $S^y$ is obtained by integrating $s_{\mathrm{av}}^{y(x)}(z)$ from the center of the slab to the vacuum region outside one of the slab's surfaces. We verified that the numerical calculations give $s_{\mathrm{av}}^{y}(z)=s_{\mathrm{av}}^{x}(z)\equiv s_{\mathrm{av}}(z) $ for $j_x=j_y$ as expected because of the systems' symmetry.\\

The calculated spin density profile for a current density equal to $10^9$ A/cm$^2$ is plotted in Fig. \ref{fig.m} for three representative systems, namely Au(001), Pt(001), and W(001). The red dots mark the position of the atoms along the $z$ direction. The surface is located at $z=z_S\sim 87$ \AA~(dashed vertical line), and the vacuum extends from there towards $z=+\infty$. At the qualitative level, $s_{\mathrm{av}}^{y}(z)$ looks quite similar in the three cases. It presents a peak at the surface atomic layer (located at $z_S\sim 87$ \AA) accompanied by an identical and opposite peak at the other slab surface (not shown).
It then decays from the surface to the interior of the slab within about four atomic layers. Inside the slab, it assumes a periodic (almost sinusoidal) behavior with the period equal to the lattice spacing. In other words, we see the formation of a spin dipole with the positive and negative polarities centered between the atoms, and integrating to zero over the bulk unit cell. CISP is therefore absent as a bulk unit cell property. It instead emerges at the surface layers because of the microscopic local gyrotropy. This behaviour is the same found for another SOC-driven effect associated to gyrotropy, namely non-Abelian diamagnetism\cite{to.08, dr.to.22}. In that case, one can calculate the effect's characteristic quantity, that is the equilibrium spin current\cite{ra.03}, and show that it is finite at surfaces, but it vanishes when integrated over the bulk unit cell \cite{dr.to.22} in the very same way as the spin density in CISP does.\\

 The quantitative comparison of the results in Fig. \ref{fig.m} shows that the surface layer's spin density of the W slab is $s_{\mathrm{av}}(z_S)\sim 10^{-4}$ $\mu_B/$\AA$^3$, that is an order of magnitude larger than that of the Au and Pt slabs, $s_{\mathrm{av}}(z_S)\sim 10^{-5}$ $\mu_B/$\AA$^3$ (note the different scale used in the three panels of Fig. \ref{fig.m}). These values can be compared to the results reported in Ref. \onlinecite{ro.ce.22} for Te. For the same current density considered here, the spin density in Te is extrapolated to be approximately $10^{-1} \mu_B$/\AA$^3$, i.e., several orders of magnitude larger than in our $5d$ metal surfaces. %of W in spite of W having a larger atomic SOC of Te (a $4d$ rather than $5d$ material). 
 The surface CISP due to local gyrotropy in an otherwise centrosymmetric system, is generally a much weaker effect than CISP in materials with bulk gyrotropic symmetry. \\
 
To see whether our estimates may depend on the surface cut, we carried additional calculations for (111) slabs. Overall, for all three materials, we find that the surface layer's spin density and the integrated surface spin density have values similar to those for the (001) case, although the spin density profile may be locally different. Once the spin density profiles are integrated, we obtain that the total spin density $S^{y(x)}$ is about $10^{4} \mu_B/$cm$^2$ for W, while it is $S^{y(x)}\approx 10^{3} \mu_B/$cm$^2$ for Au and Pt. These are small values, but, as stated in Ref. \onlinecite{to.kr.15}, they could in principle be measured in optical experiments with the current resolution. \\

%Overall, our results confirm the general viewpoint that W is one of the best materials for spin-charge conversion in general, and for surface CISP in particular. In contrast, the finding that Pt(001) and Au(001) display a similar spin density is somewhat surprising. Pt is well established as the best $5d$ metal for SHE,  while SHE is instead negligible in Au (see following). Based on this fact, it is often assumed in the literature that Pt should similarly be the best material for CISP. Our results suggest that this is not the case. SHE and CISP are different effects, the first being a bulk effect, while the second being caused by states scattering off the surface and related to gyrotropy. As such, there may be no quantitative correlation between the two. \\         
\begin{figure}[h]
\centering\includegraphics[width=0.48\textwidth,clip=true]{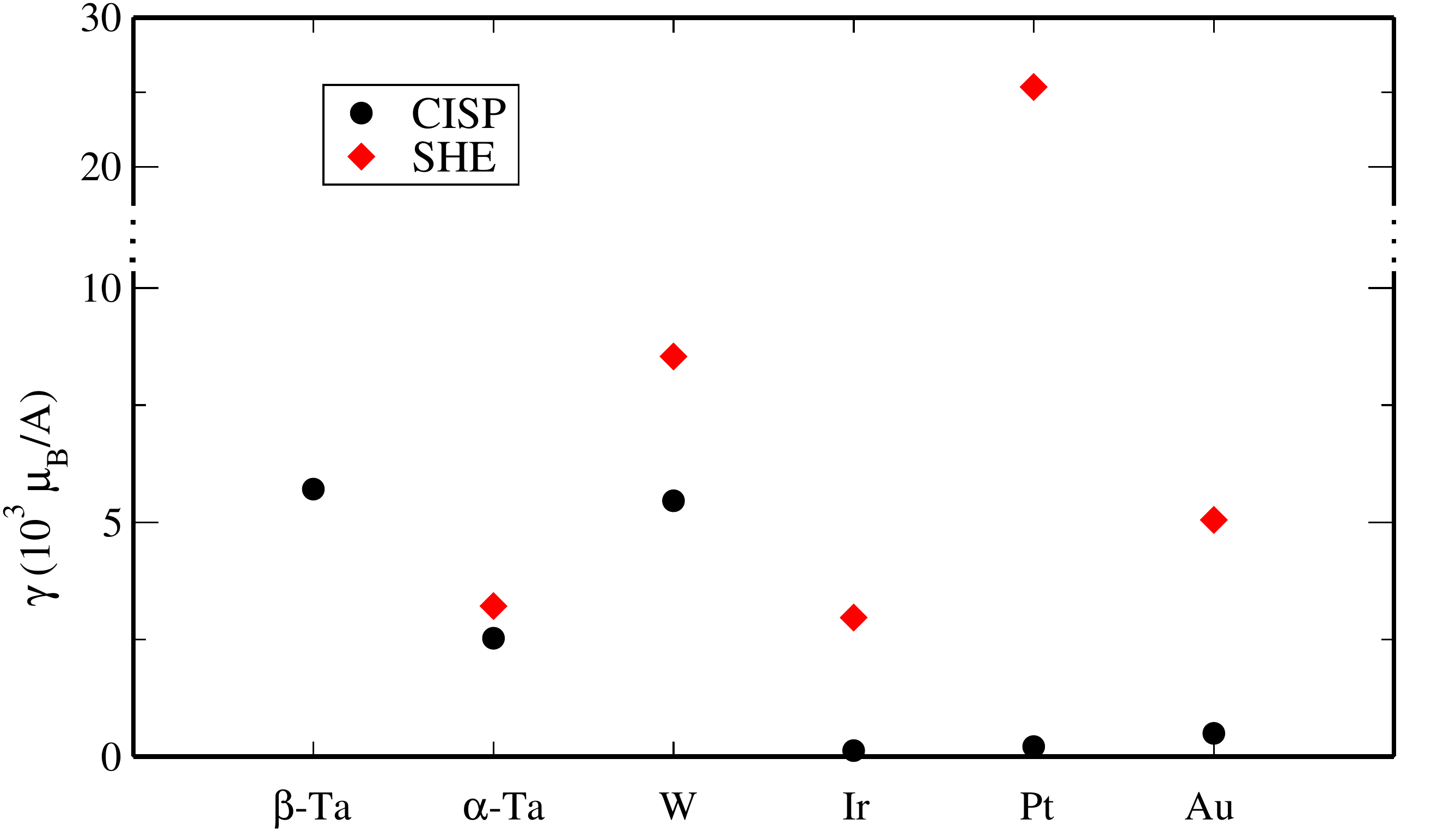}
\caption{Surface CISP (SHE) factor $\gamma$ ($\gamma_{SHE}$) for the various investigated $5d$ metals.  }
\label{fig.m_element}
\end{figure}

A systematic comparison of the surface CISP across the various considered $5d$ materials is done by plotting in Fig. \ref{fig.m_element} the parameter $\gamma\equiv \gamma^x_{y}=\gamma^y_{x}$ of Eq. (\ref{eq.SvsJ}) (see the black dots). %The black dots represent the results for the room temperature stable crystal structures, i.e. fcc (Au, Pt, Ir), bcc (Ta, W), and hcp (Hf, Re, Os). 
%For anisotropic hcp surfaces we show both $\gamma_y^x$ and $\gamma_x^y$.
In the case of Ta, we also consider the so-called $\beta$-phase, which is commonly regarded as a high performing material for spin-charge conversion \cite{li.ch.12,sa.om.18}. %Besides, we also carried out calculations assuming that all materials had fcc structures, and the results for the fcc Hf, Ta, W, Re, and Os slabs are represented by the red squares. 
The most striking conclusion that can be drawn based on the results, is that the surface CISP is about an order of magnitude larger for Ta and W, than for the late $5d$ materials. In fact, $\gamma$ is about $10^3\mu_B/$A for Ta and W, while it is about $10^2\mu_B/$A for Pt and Au. This indicates that there might be a correlation between CISP and the Hund's coupling, although we must stress that CISP depends on the fine details of the slabs' intricate band structure and on the filling of the electronic states according to the out-of-equilibrium electron distribution. A simple interpretation of the results based on a single atomic property is therefore not possible. %Extracting simple qualitative argument to predict the magnitude of CISP is not possible. 
Yet, looking at our calculations, we can suggest that spintronic devices aiming at exploiting surface CISP, should be made either of Ta or of W films for achieving the best possible performances. \\ 

%$\gamma$ (unit $10^{-3}$ $\mu_B$/A)
%Hf: 1.78, 2.53\\
%$\beta$ Ta: 5.7\\
%Ta: 2.52\\
%W: 5.46\\
%Re: 0.29, 0.99\\
%Os: 0.059, 3.13\\
%Ir: 0.13\\
%Pt: 0.21\\
%Au: 0.49\\

Finally, we turn to the question about the relative importance of CISP and of the SHE in $5d$ metallic films. To do so, we need to quantify the spin accumulation induced by the SHE and compare it to the spin density due to CISP. We then consider a semi-infinite system along the $z$ direction, with the surface located at a position $z=z_S$, and vacuum extending from $z_S$ towards $z=+\infty$. The charge current $j_x$ flows along the $x$ direction as in the CISP calculations above. Assuming diffusive spin dynamics, the SHE induces a $z$-dependent spin current and a spin density that are given by the equations\cite{DyaKha2017,bo.to.19} 
\begin{eqnarray}
j^y_{z}(z)=[\theta_{SHE}j_x-D \partial_z s_{SHE}^y(z)],\label{eq.SHE_current}  \\
\partial_z^2 s_{SHE}^y(z)-\frac{1}{l_{SD}^2}s_{SHE}^y(z)=0,\label{eq.SHE_spin}
\end{eqnarray}
where $\theta_{SHE}$ is the spin Hall angle, $l_{SD}$ is the spin diffusion length, and $D$ is the diffusion constant. 
Eq. (\ref{eq.SHE_spin}) implies that the spin-density vanishes exponentially from the surface towards the bulk of the system, and the solution reads
\begin{equation}
    s_{SHE}^y(z)=s e^{-(z_S-z)/l_{SD}},\,\,\,\, z\leq z_S,
\end{equation}
where $s$ is the spin density at the surface, i.e. at $z=z_S$.
Eq. (\ref{eq.SHE_current}) must be accompanied by the boundary condition $j^y_z\vert_{z=z_S}=0$, which enforces 
the spin current to vanish at the surface as it can not flow into the vacuum region. Thus, the integrated spin density is
%\begin{equation}
%S^y=\int_0^\infty dz s^y(z)=-\frac{\theta_{SHE}}{D}l_{SD}^2j_x,
%\end{equation}
%which can be rewritten as
\begin{equation}
S^y_{SHE}=\gamma_{SHE}j_x   
\end{equation}
and defines the spin accumulation. The parameter
$\gamma_{SHE}=\theta_{SHE}l_{SD}^2/D$ expresses the magnitude of the spin density induced only by the SHE as opposed to $\gamma$ in Eq. (\ref{eq.SvsJ}) that uniquely characterizes CISP. $\gamma_{SHE}$ can be directly computed provided that $\theta_{SHE}$, $l_{SD}$, and $D$ are known to a reasonable accuracy. Unfortunately, however, the reported values for these quantities vary widely from experiment to experiment \cite{si.va.15,ho.13} likely because of the different crystalline quality of the measured samples and unavoidable extrinsic contributions from disorder or impurities \cite{sa.ya.16}. We therefore employ here the theoretical estimates provided by Nair {\it et al.} in Ref. \onlinecite{na.ba.21}, which are based on scattering theory combined with DFT. The calculated values of $\gamma_{SHE}$ are presented in Fig. \ref{fig.m_element} as red squares. No data are available for $\beta$-Ta. \\ %In case of hcp materials we report the results obtained with data referring to the charge current parallel and perpendicular direction with respect to the hcp hexagonal axis\cite{na.ba.21}.\\

% $\gamma_{SHE}\approx 2.5 \times 10^{4}\mu_B$\ A is much larger for Pt than for any other material because the spin Hall angle of Pt is the largest ($\theta=4.02\%$). % For Au, $\gamma_{SHE}$ is not vanishing in spite of the negligible spin Hall angle ($\theta_{SHE}=0.25\%$) because of the huge spin diffusion length $l_{SD}=50.9$ nm.  
In Ta and W, $\gamma_{SHE}$ is almost as large as the CISP factor $\gamma$. This means that the integrated spin density induced by the SHE and CISP are comparable in these materials. However, it is important to point out the different length scale over which the spin density is localized. In CISP, the spin density is mostly confined within few atomic layers around the surface (see Fig. \ref{fig.m}), i.e., over a length $l_{CISP} \approx 1$ nm. In contrast in the SHE, the spin density is accumulated over a characteristic length dictated by the spin diffusion length. In case of W, $l_{SD}$ is $29.6$ nm (about 140 atomic layers). This means that, at the surface layer, $s_{SHE}/s_{CISP}\approx l_{CISP}/l_{SD}\approx 3 \times 10^{-2}$. The spin density originating from CISP is two orders of magnitude larger than that due to the SHE, or, in other words, the contribution of the SHE in W is negligible compared to that of CISP. 
Similar considerations also apply to Ta, although we obtain $s_{SHE}/s_{CISP}\approx 10^{-1}$ since $l_{SD}$ is much smaller ($6.16$ nm) than in W. Nonetheless, we must point out that the spin Hall angles of W and Ta taken from Ref. \onlinecite{na.ba.21} are very small, i.e. $\theta_{SHE}\approx 0.4$ and $0.5$. While the value for Ta appears in fair agreement with the experimental results \cite{sa.om.18}, in the case of W, the experimental estimates for $\theta_{SHE}$ span from $0$ to $5\%$ (Refs. \onlinecite{qu.hu.14,fr.wi.18}). A theoretical evaluation of the intrinsic contribution based on the Berry curvature of the band structure gave $2\%$ (Ref. \onlinecite{su.wa.17}). Using this value, we would obtain a slight increase in the relevance of the SHE.\\

In the case of Au, $\gamma_{SHE}$ is not vanishing in spite of the negligible spin Hall angle ($\theta_{SHE}=0.25\%$). This is because of the huge spin diffusion length, $l_{SD}=50.9$ nm. As a result, we find that $s_{SHE}/s_{CISP}\approx 0.5$. Thus, in Au, the SHE and CISP contributions to the surface spin density are somewhat comparable.   
Finally, in Pt, we find that 
$\gamma_{SHE}\approx 2.5 \times 10^4$ is much larger than in any other material because of the short spin diffusion length $l_{SD}=5.21$ nm and of the large spin Hall angle ($\theta=4.02\%$). %When compared with the magnitude of the CISP, $\gamma\approx 2 \times 10^2$, because of the short spin diffusion length $l_{SD}=5.21$nm and of the large spin Hall angle ($\theta=4.02\%$). 
Consequently, at the surface layer we obtain $s_{SHE}/s_{CISP}\approx 25$. This means that the surface spin density is mostly induced by the SHE rather than by CISP. Pt represents the opposite case of W. Some important experiments with Pt, where CISP due to the surface states was considered as the dominant contribution to the surface spin density \cite{mi.mo.11}, might need to be revisited in terms of the SHE.\\
    
Recently, spin-polarized positron beam experiments were performed to directly probe CISP at some $5d$ metal surfaces \cite{zh.ya.14}. To compare our calculations with these experimental results, we define the spin-polarization (SP) as the spin density divided by the charge density in proximity to the surface layer. The largest calculated SP is obtained for W and is equal to about $10^{-5}$ for a charge current density as large as in experiments ($10^{5}$ A/cm$^2$). This SP is several orders of magnitude smaller than the experimentally reported one, that is about $0.1$. Overall, all experimental values appear surprisingly large, and even much larger than the results in Refs. \onlinecite{ca.su.22, ro.ce.22} for bulk Te, which, as mentioned above, is a ``strong'' gyrotropic material and is expected to show enhanced CISP compared to any metallic surfaces. Resolving such disagreement between the calculated and measured SP will represent an important issue for potential future studies.\\

In conclusion, we provided an estimate of the CISP magnitude for several $5d$ metallic surfaces. The W surface shows the largest effect, while CISP is an order of magnitude smaller at Pt and Au surfaces. We also showed that the spin density due to CISP may often be comparable to the spin accumulation induced by the SHE. However, the two effects appear at different length scales. In the case of W, the spin density at a film's surface layer is mostly due to CISP. In contrast, in the case of Pt, the surface spin density is caused mostly by the SHE. These observations may be valuable for the interpretation of experiments.\\

{\it Acknowledgments.} A.D. was supported by Science Foundation Ireland and the Royal Society through the University Research Fellowships URF-R1-191769, and by the European Commission through the H2020-EU.1.2.1 FET-Open project INTERFAST (project ID 965046). I.V.T. acknowledges support by Grupos Consolidados UPV/EHU del Gobierno Vasco (Grant IT1453-22) and by the grant PID2020-112811GB-I00 funded by MCIN/AEI/10.13039/501100011033. The computational resources were provided by Trinity College Dublin Research IT. 
\bibliography{ref}

\end{document}